\documentclass[a4paper]{jpconf}
\usepackage{graphicx}
\usepackage{amssymb}
\begin{document}
\title{Finite conformal quantum gravity and \\
spacetime singularities}

\author{Leonardo Modesto and Les\l{}aw Rachwa\l{}}

\address{Department of Physics, Southern University of Science and Technology,\\
 Shenzhen 518055, P. R. China}

\ead{lmodesto@sustc.edu.cn}
\vspace{-0.2cm}
\ead{grzerach@gmail.com}

\begin{abstract}
We show that a class of finite quantum non-local gravitational theories is conformally invariant at classical as well as at quantum level. This is actually a range of conformal anomaly-free theories in the spontaneously broken phase of the Weyl symmetry. At classical level we show how the Weyl conformal invariance is able to tame all the spacetime singularities that plague not only Einstein gravity, but also local and weakly non-local higher derivative theories. The latter statement is proved by a singularity theorem that applies to a large class of weakly non-local theories. Therefore, we are entitled to look for a solution of the spacetime singularity puzzle in a missed symmetry of nature, namely the Weyl conformal symmetry. Following the seminal paper by Narlikar and Kembhavi, we provide an explicit construction of singularity-free black hole exact solutions in a class of conformally invariant theories.
\end{abstract}

\section{Introduction}

\hspace{0.4cm} There is one old symmetry that has always played a very crucial role in fundamental physics. This is the conformal symmetry, which is enjoyed by physical systems whose description does not depend on any characteristic scale.  
Historically this symmetry dates back to Hermann Weyl who discovered the concept of local gauge symmetries. At first, the gauge symmetries were applied to the local change of scale of the length measuring rulers. Since the measurements of space (and time) distances are the operational basis of general relativity (GR), we can understand why the conformal symmetry as a gauge symmetry first appeared in the gravitational setup. Only later the ideas of gauge symmetries were used in the non-gravitational setting of matter field theories, namely in the particle physics and condensed matter. Moreover, the conformal symmetry was found back in gauge theories too but the necessary condition here was to be in four spacetime dimensions. Indeed, the Maxwell's and Yang-Mills theories revealed to be exactly conformally invariant at the classical level. In the classical notion of conformal invariance the spacetime metric undergoes a rescaling by some always positive scalar function eventually dependent on the spacetime points. This is also the meaning of conformal symmetry used in classical GR, where it has been very useful to study the black holes' near-horizon geometry, the FRW cosmologies, and it was also a crucial element for the construction of the Newman-Penrose formalism.

There is also a different meaning of the same symmetry having to do with some properties of correlation functions on a given spacetime background. First of all, the change of scale on flat spacetime can be very easily transmitted to a change of coordinates and then the symmetry shows up in a somehow global disguise. One can then study its extension to the full conformal group, which on a flat 4-dimensional Minkowski spacetime acts as a realization of the SO(2,3) group. This global symmetry initially finds its manifestation in some conformally invariant classical field theory models, but its true power is in constraining the correlation functions at the quantum level. Very soon, it was proven that the requirement of scale invariance at the quantum level implies that the beta functions of the couplings must exactly vanish. Moreover, the scale-invariance must be promoted to conformal symmetry to have a fully consistent field theory. The correlators of such very special theory show amazing symmetries that are described by the conformal algebra and the theories themselves are called conformal field theories (CFT's). When the beta functions are zero, then also the UV-divergences of the theory are absent, so the theory in question is UV-finite. There are very special examples of such theories and one of the most (super-) symmetric example is ${\cal N}=4$ Super-Yang-Mills theory in four dimensions. This line of development found its culmination in the AdS/CFT duality, where the ideas of CFT were also used to study the gravitational physics on the anti-de Sitter background and vice versa.

A natural question is whether these two ways of using conformal symmetry may be joined together. Since the conformal transformations are of spacetime character (not in the internal group of matter symmetries) then the merging shoukd necessarily involve gravity. At the beginning, the crucial issue was how to make finite quantum gravity. Unfortunately, the quantum gravitational theories were from the beginning beset by UV-divergences and the first theory of gravity invented by Einstein in 1915 revealed to be non-renormalizable at the quantum level. There was completely no control over divergences, and so the theory had to be modified. The addition of terms with higher derivatives in the action helps with the renormalizability and even one-loop super-renormalizability was achieved \cite{shapiro3}-\cite{modesto4}. In this way a much bigger control over divergences was gained. Additionally the inclusion of non-locality solved the unitarity issue. The last piece in this quest for UV-completion required the construction of a theory where all divergences were precisely cancelled. This was achieved for the first time by the authors in \cite{finite, universality, fingauge}. This result opened the possibility to have conformally invariant theories also at quantum level because the condition for the cancellation of the conformal anomaly is the vanishing of beta functions for all couplings. Now, we also see that the implementation of conformal symmetry in a fully gravitational and fully quantum framework has very profound and unexpected consequences. Indeed, not only we have a theory free of pathologies in the UV, but also we have exact spacetime solutions that are without singularities. This was another problem that hampered Einstein's GR, but now due to conformal quantum gravity we can firmly predict and describe what is beyond the ``hypothetical'' singularity at the black hole origin, Big Bang event, or in many others situations, which in the past created problems with predictability and extendability of the classical Cauchy evolution. The ideas of using conformal symmetry for solving these problems originated from the seminal work by Narlikar and Kembhavi \cite{Narlikar}.

Below we describe in some details the explicit construction of the finite quantum gravity with its natural conformal extension, and the mechanism for the resolution of spacetime singularities.

\section{UV-finite quantum gravity}

\hspace{0.4cm} As motivated in the introduction Einstein's gravity must be modified to achieve UV-finiteness. We here consider a general gravitational theory described by the following Lagrangian density
\begin{equation}
\mathcal{L}_{\rm g} = -  2 \kappa_{4}^{-2} \, \sqrt{|g|} 
\left( { R} +R \gamma_0(\Box)R+R_{\mu\nu} \gamma_2(\Box)R^{\mu\nu}+R_{\mu\nu\rho\sigma} \gamma_4(\Box)R^{\mu\nu\rho\sigma}
+ {\cal V}({\cal R}) 
\right),
\label{gravityG}  
\end{equation}

where the first two form-factors are defined in terms of entire functions $H_0$, $H_2$ by
\begin{equation}
\gamma_0(\Box)=-\frac{2\left(e^{H_0}-1\right)+4\left(e^{H_2}-1\right)}{12\Box}+\gamma_4(\Box)\quad{\rm and}\quad\gamma_2(\Box)=\frac{e^{H_2}-1}{\Box}-4\gamma_4(\Box)\,.
\end{equation}
We remark that the third form-factor $\gamma_4(\Box)$ remains arbitrary and is constrained only by renormalizability properties of the theory. The functions $H_0$, $H_2$ are required to be analytic entire functions on the whole complex plane without poles and with asymptotic polynomial behaviour. The
$\gamma_4$ is an analytic function without poles and with asymptotic behaviour compatible with the renormalizability of the theory. In $\cal V(R)$ we collect terms, which are cubic or higher powers in generalized gravitational curvatures $\cal R$. These terms are crucial in making the theory finite as we will show below.

The theory described by the Lagrangian (\ref{gravityG}) has the same symmetry as Einstein's gravity (diffeomorphism invariance). However, it is non-local due to the non-polynomial, but still analytic, structure in the form factors being functions of the covariant box operator \cite{review, Calcagni:2010ab}. The two entire functions $H_0$ and $H_2$ fully specify the structure of the propagator for the gravitational fluctuations. If they satisfied the conditions stipulated above there are only two propagating degrees of freedom, namely the same as the two polarizations of massless gravitons in quantum Einstein's gravity. The theory is also unitary at perturbative level in Minkowski space.

For questions about renormalizability and UV divergences we need to study the behaviour of the theory for large values of momenta. We decided to concentrate on the theory, which in the UV regime tends to a polynomial higher derivative theory. This is achieved by requiring that for large values of the argument around real axis the form-factors $\gamma_i(\Box)$ for $i=0,2$ or 4 have monomial asymptotics. The first examples of such form-factors were given in the seminal papers by Kuz'min and Tomboulis \cite{Tombo1,Tombo2, kuzmin}. For simplicity we assume that the exponent there is a natural number $\gamma$ the same for all form-factors. It was found that for $\gamma\geqslant3$ the theory is one-loop superrenormalizable \cite{modesto1, modesto2, modesto4}, which means that only one-loop divergences survive and the theory from two loops on is completely convergent.

Having a super-renormalizable theory it is easy to look for an extension, in which all one-loop divergences are absent. For this purpose we use the terms in $\cal V(R)$. One can convince himself, using for example the covariant Barvinsky-Vilkovisky (BV) formalism for the quantum effective action, that the divergences can be cancelled by a precise choice of the front coefficients for the following operators, 
\begin{equation}
s_1 R^2 \Box^{\gamma-2} R^2,
\quad 
s_2 R_{\mu\nu}^2 \Box^{\gamma-2} R_{\rho\sigma}^2,
\quad 
s_3 R_{\mu\nu\rho\sigma}^2 \Box^{\gamma-2}R_{\kappa\lambda\tau\omega}^2
\end{equation}
Indeed, in four dimensions (and for asymptotically monomial form-factors) we have to make zero three beta functions for the dimensionless couplings in front of the operators quadratic in the gravitational curvatures, namely 
\begin{equation}
\beta_{R^2},\quad
\beta_{R_{\mu\nu}^2},
\quad \beta_{R_{\mu\nu\rho\sigma}^2}
\label{killers123}
\end{equation}
Actually due to a topological relation, using Gauss-Bonnet density, these beta functions are not independent (see also \cite{betaGN}). However, for full conformal invariance we have to secure that there are no divergent terms also at the boundary of our spacetime. It turns out that we have to cancel the fourth beta function 
\begin{equation}
\beta_{\Box R} \quad \mbox{by using operators of the type} \quad s_4(\Box R)\Box^{\gamma-2}R^2 . 
\label{killer4}
\end{equation}
By using dimensional arguments, counting powers of covariant curvatures, and the power of BV formalism we see that the contribution of the listed above terms (\ref{killers123}), (\ref{killer4}) in $\cal V(R)$ to the corresponding beta function is linear in their coefficients $s_i$ ($i=1,2,3,4$). The matrix giving such linear mapping can be easily shown to be non-degenerate hence the system of the equations for cancellation of these beta functions has always real solutions for the coefficients $s_i$. This construction bring us a UV-finite quantum gravitational theory.

\section{Finite conformal quantum gravity}

\hspace{0.4cm} 
When we have a theory that behaves well at quantum level (namely does not have any divergences) we may compute its predictions to whatever accuracy with special care of finite terms. But computing these observables we notice that the theory possesses some hidden symmetry, which is revealed for example in the form of scattering amplitudes for on-shell gravitons \cite{scattering}. This hidden symmetry is the Weyl's conformal invariance, and the main effort now is only to make such symmetry explicit in the action and on every step of the computation. In other words, we want to rewrite our finite quantum gravity in a conformally invariant way. We here use the compensating field method, which can be employed in the Lagrangian density (\ref{gravityG}) making the replacement $g^{\mu\nu}= \left( \phi^2 \kappa_4^2 \right)^{-1} \hat{g}^{\mu\nu}$, namely 
\begin{eqnarray} 
&& 
\hspace{-0.5cm} 
\mathcal{L}_{\rm conf.} = -  
2  \, \sqrt{ |\hat g| } 
\left[   \phi^2 R(\hat{g}) +
6 \hat{g}^{\mu\nu} (\partial_\mu \phi) (\partial_\nu \phi) \right]
\label{Conf2}\\  
&& \hspace{-0.5cm} 
  - \frac{  2 }{\kappa_{4}^{2}}   \sqrt{|g|} \left[ 
{ R}({g}) \, 
 \gamma_0( {\Box} )
 { R} (g)
 + R_{\mu\nu} (g) \, 
\gamma_2( {\Box} )
 R^{\mu\nu} (g) 
 + R_{\mu\nu\rho\sigma} (g) \, 
\gamma_4( {\Box} )
 R^{\mu\nu\rho\sigma} (g) 
+ { {\cal V}({\cal R}(g))} \, 
\right] \Big|_{g=\phi \hat{g}} \, \,  , \nonumber
\end{eqnarray}
where by $\big|_{\phi \hat{g}}$ we mean that the metric $g^{\mu\nu}$ must be replaced with a product
$\left( \phi^2 \kappa_4^2 \right)^{-1} \hat{g}^{\mu\nu}$. 

In (\ref{Conf2}) we have an extra ghost-like degree of freedom. However, it can be eliminated using the extra symmetry of the theory, namely conformal invariance. Indeed, in this theory we have a dynamical metric tensor $\hat g_{\mu\nu}$ and a scalar field $\phi$. When the unitary gauge $\phi = {\rm const.}$ is enplaned then we can come back to our finite quantum theory.
Moreover, the quantum properties of the theory can not change in different gauges, and the theory must be UV-finite also when the scalar field is not completely gauged away, but another gauge is implemented.

We notice that the metric $g^{\mu\nu}=\left( \phi^2 \kappa_4^2 \right)^{-1} \hat{g}^{\mu\nu}$ (and therefore the action  (\ref{Conf2})) is invariant under the conformal rescalings
\begin{equation}
\hat{g}_{\mu\nu} \rightarrow \Omega^2(x) \, \hat{g}_{\mu\nu} \, , 
\quad \phi \rightarrow \Omega^{-1}(x) \, \phi ,
\label{conftransf}
\end{equation}
meaning that $\phi$ describes clocks and rulers (physical scales) 
while $\hat{g}_{\mu\nu}$ contains exclusively information 
about the light-cone causal structure of the spacetime. 
After gauge fixing, $\phi$ can be just a constant and the 
distances are now measured by the metric $\hat{g}_{\mu\nu}$ 
or equivalently by the metric $g_{\mu\nu}$. The two metrics are now identified and, therefore,
they have the same number of degrees of freedom, namely $10$.  By choosing a conformal gauge we select a vacuum for the field $\phi$ hence this process can be viewed as a spontaneous symmetry breaking. In other words the ``effective number of d.o.f." is actually $10$ before and after breaking of conformal invariance. The conformal ``Higgs" mechanism here consists of moving the d.o.f., measuring time and space, from $\phi$ to the metric $\hat{g}_{\mu\nu}$. Such degree of freedom is the analog of the Higgs field giving mass to particles in the standard model of particle physics (SM). 

The theory defined by the Lagrangian (\ref{Conf2}) can be analyzed like any other QFT. One can study the UV divergences of the theory when both the scalar and tensor perturbations propagate in any conformal gauge. We consider the fluctuations of the metric 
$\hat{g}_{\mu\nu}$ and of the scalar field $\phi$. Afterwords, we expand the action around a given spacetime background and a given background for the scalar field. The strict power counting and one-loop super-renormalizability of this theory can be easily shown, for more details we refer to \cite{conf}. Moreover, we have proved that the theory is unitary and all divergences are cancelled due to terms in $\cal V(R)$ regardless of the chosen conformal gauge in the quantization process. We also remind that the Fujikawa integration measure in path integrals is naturally conformally invariant. In the dimensional regularization scheme (DIMREG) some special care must be exerted when the theory is defined away from $d=4$ spacetime dimensions. It is a known fact that the only way how conformal anomaly can arise is due to non-vanishing beta functions. If we do not have divergences, then we are also safe against conformal anomaly. This claim is independent whether the vacuum is in the broken or unbroken phase of the symmetry.

We showed that the theory is UV-finite (no divergences) and free of conformal anomaly (all the beta functions vanish), therefore, conformal invariance is preserved at the quantum level. The classical action of the theory (\ref{Conf2}) is also manifestly invariant, hence we have found a class of conformally invariant finite quantum gravity theories. These theories are in the spontaneously broken phase of the quantum conformal Weyl symmetry. 

\section{Resolution of the black hole singularity in conformal gravity}

\hspace{0.4cm} We propose a solution of the black hole singularity problem based on conformal symmetry. This idea was already explored in the 70's by Narlikar and Kembhavi \cite{Narlikar}, but we here reconsider and generalize their ideas for two reasons. First, we can use conformal methods because we have conformal quantum gravity, while they used it only at the classical level for a conformally invariant theory. Secondly, we made this idea to work not only for Big Bang singularities of FRW spacetimes (when the use of conformal methods seems rather trivial, cf. \cite{cosmo}), but foremost for black holes. Moreover, we can also get rid off of other pathologies like closed time-like curves, which by conformal rescalings can be reduced to have zero time lapse. Therefore, we conclude that we can make any singularity appearing in exact solutions of GR innocuous by a proper conformal rescaling.

Spacetime singularities are properties of some exact solutions of Einstein's gravity, where the group of symmetry is the group of diffeomorphisms. We remark that, contrary to some previous claims, theories with weak non-locality are also not able to get rid out of the spacetime singularities \cite{exactsol, Calcagni, Modesto:2010uh}. Weak non-locality is not enough and we need to enlarge the underlying diffeomorphism (Diff.) invariance of general relativity to include the Weyl conformal invariance. Indeed, only an enhanced symmetry, here local conformal symmetry is uniquely able to tame the singularities that plague not only Einstein gravity, but also local and weakly non-local higher derivative theories. We show this explicitly by analyzing the exact equations of motion (EOM).

We start with a Riemannian spacetime manifold $\mathcal{M}$ with a metric tensor $\hat{g}_{\mu\nu}$ and a scalar field $\phi$ as discussed in the previous section. 
Then the EOM for the metric $\hat{g}_{\mu\nu}$ in the theory (\ref{Conf2}) with the choice $\gamma_4=0$ read: 
\begin{eqnarray} 
&&  \hspace{-0.2cm}
  \phi^2 \hat{G}_{\mu\nu}  =
   \nabla_\nu \partial_\mu \phi^2 - \hat{g}_{\mu\nu} \hat{\Box} \phi^2 
 -  6 \left( (\partial_\mu \phi) (\partial_\nu \phi) - \frac{1}{2} \hat{g}_{\mu\nu} g^{\alpha \beta}
 (\partial_\alpha \phi) (\partial_\beta \phi) \right) \nonumber \\
 && 
\hspace{1.5cm}  - \frac{ \delta  \left(  \sqrt{|g |}\left( 
  R 
\gamma_0(\Box) R+R_{\mu\nu} 
\gamma_2(\Box) R^{\mu\nu} + {\cal V}\right)
\right) \Big|_{\phi \hat{g}} 
}{ \sqrt{| g_{\phi \hat{g}}  |} \, \delta \hat{g}^{\mu\nu} }  ,
 \label{EOMg}
  \end{eqnarray}
\vspace{-0.5cm}
while the EOM for $\phi$ is:
\begin{equation}
  \hat{\Box} \phi = \frac{1}{6} \hat{R} \phi 
 - \frac{ 1 }{\sqrt{| g_{\phi \hat{g}}  |} } 
 \frac{ \delta  \left(  \sqrt{|g |} 
 \left(
  R 
\gamma_0(\Box) R+R_{\mu\nu} 
\gamma_2(\Box) R^{\mu\nu} + {\cal V}
\right)\right) \Big|_{\phi \hat{g}} 
}{ \delta \phi } \, .
\label{EOMp}
\end{equation}
We can choose the operators in $\cal V(R)$ to be function of  Ricci tensors only and not of the Riemann
tensor. Since then all the operators resulting from the variation are at least linear in the Ricci tensor $R_{\mu\nu}$
(note that we here mean the Ricci tensor without a hat), then the Schwarzschild metric is 
an exact solution of the conformally invariant theory (\ref{Conf2}). This can be easily
verified replacing the solution  $\hat{g}_{\mu\nu} = g_{\mu\nu}^{{\rm Sch}}$ and $\phi = \kappa_4^{-1}$
in the above EOM (\ref{EOMg}) and (\ref{EOMp}). 
We notice as the result of the theorem that if the theory is general higher derivative or non-local then the same applies and we find Schwarzschild (with its singularity at the origin) as the exact solution of the theory \cite{exactsol}.

However, the EOM are conformally invariant, 
hence if we consider another manifold $\mathcal{M}^*$ obtained from $\mathcal{M}$ by a conformal transformation  (\ref{conftransf}) to $^*$ quantities, then also $\hat{g}_{\mu\nu}^{*}$ and $\phi^{*}$ satisfy the EOM (\ref{EOMg}) and (\ref{EOMp}).
The transformation $\phi \rightarrow \phi^*$ is mathematically valid provided $\Omega^{-1}$ does not vanish (or become infinite). It is assumed that $\Omega=\Omega(x)$ is a twice differentiable positive function (or a smooth function in a non-local theory) 
of the spacetime coordinates $x$.

Starting from the exact Schwarzschild solution we can now construct an infinite class of other exact solutions, all of them conformally equivalent to the Schwarzschild metric, by doing rescalings (\ref{conftransf}) to both the scalar field $\phi$ and the metric $\hat g_{\mu\nu}$ with some $\Omega$.
%
 For $\phi= \kappa_4^{-1}$ and $\Omega=1$ we get the Schwarzschild spacetime 
 Moreover, 
$ \hat{R}_{\mu\nu}(\hat{g}^*_{\mu\nu}) \neq 0$,
so the conformally rescaled Schwarzschild metric is not any more Ricci-flat.
 
In this large class of solutions we can pick out a subclass of singularity-free black hole spacetimes. One example of which is provided by the following singularity-free exact black hole solution (in any conformally invariant theory), obtained by rescaling the Schwarzschild metric by a suitable $\Omega$:
\begin{equation}
ds^{* 2} =  \left( 1+ \frac{L^4}{r^4} \right)\left[ \left( 1- \frac{2 m}{r} \right) dt^2 + \frac{dr^2}{1- \frac{2 m}{r}} + r^2 d \Omega^{(2)} \right] , \label{NRBH} \quad \phi^* =  \left( 1+ \frac{L^4}{r^4} \right)^{-1/2} \kappa_4^{-1} \, . 
\label{solex}
\end{equation}
If this metric is considered in GR, then one can easily show that the spacetime is regular at $r=0$. This is supported by the regularity of all curvature invariants (like Kretschmann scalar) and by the geodesic completion for massive as well as massless particles moving in this geometry \cite{completeness}. In the solution (\ref{solex}) $L$ is an arbitrary length scale. This can be naturally chosen either to be the Planck length or the mass of the black hole $m$. The warp factor $\Omega(x)$ chosen in (\ref{solex}) is one of the infinitely many possible which render the transformed spacetime completely regular and without any singularities. An analogous rescalings can be found to make regular Kerr-Newman rotating and charged black hole backgrounds. In a conformally invariant theory this metric turns out to be physical and observable if the conformal symmetry is spontaneously broken. 

In conformal gravity the black hole thermodynamics and the Hawking evaporation process had been  presented in \cite{evaporation}. Similarly, the issue with the black hole entanglement entropy in the quantum field theory framework had been solved in \cite{entanglement}. Basically, the black hole entropy is finite because the theory (\ref{gravityG}) is UV-finite at the quantum level.


\section*{References}
\begin{thebibliography}{99}

\bibitem{shapiro3} Asorey M,  L\'opez J L , Shapiro I L,
{\it Int.\ J.\ Mod.\ Phys.} A 12 5711 (1997) ({\it Preprint} hep-th/9610006)

\bibitem{Tombo1} Tomboulis E T, 
({\it Preprint} hep-th/9702146)

\bibitem{Tombo2}  Tomboulis E T,
 {\it  Mod.\ Phys.\ Lett.} A {\bf 30}, no. 03n04, 1540005 (2015)

\bibitem{Khoury} Khoury J,
{\it  Phys.\ Rev.} D {\bf 76}, 123513 (2007) ({\it Preprint} hep-th/0612052)

\bibitem{modesto1} Modesto L,
{\it  Phys. \ Rev.} D {\bf 86}, 044005 (2012) ({\it Preprint} hep-th/1107.2403) 

\bibitem{modesto2} Modesto L,
 {\it Astron. Rev.} 8.2 (2013) 4-33 ({\it Preprint} hep-th/1202.3151)

\bibitem{modesto4} Modesto L,
{\it Preprint} hep-th/1202.0008

\bibitem{finite} Modesto L and Rachwal L, 
  {\it Nucl.\ Phys.} B {\bf 889}, 228 (2014) ({\it Preprint} hep-th/1407.8036)

\bibitem{universality} Modesto L and Rachwal L, 
  {\it Nucl.\ Phys.}  B {\bf 900}, 147 (2015) ({\it Preprint} hep-th/1503.00261)

\bibitem{fingauge} Modesto L, Piva M and Rachwal L, 
{\it Phys.\ Rev.}  D {\bf 94}, no. 2, 025021 (2016) ({\it Preprint} hep-th/1506.06227)

\bibitem{Narlikar} 
  Narlikar J V and Kembhavi A K,
 {\it Lett.\ Nuovo Cim.}  {\bf 19}, 517 (1977)

\bibitem{kuzmin} Kuz'min Y V, 
{\it Yad.\ Fiz.} \ {\bf 50}, 1630 (1989) [{\it Sov.\ J.\ Nucl.\ Phys.}\ {\bf 50}, 1011 (1989)].

\bibitem{review} Modesto L and Rachwal L, {\it Int.\ J.\ Mod.\ Phys.}  D {\bf 26}, 1730020 (2017)

\bibitem{Calcagni:2010ab} Calcagni G and Nardelli G,  
{\it Phys.\ Rev.}  D {\bf 82}, 123518 (2010) ({\it Preprint} hep-th/1004.5144)

\bibitem{betaGN} Modesto L, Rachwal L and Shapiro I L, 
{\it Preprint} hep-th/1704.03988

\bibitem{scattering} Dona P, Giaccari S,  Modesto L, Rachwal L and Zhu Y, 
{\it J. High Energy Phys.} JHEP08(2015)038 

\bibitem{conf} Modesto L and Rachwal L, 
{\it Preprint} hep-th/1605.04173

\bibitem{exactsol} Modesto L and Rachwal L, 
{\it J. High Energy Phys.} JHEP1512(2015)173 ({\it Preprint} hep-th/1506.08619)

\bibitem{cosmo}  Koshelev A S, Modesto L, Rachwal L and Starobinsky A A, 
{\it J. High Energy Phys.} JHEP1611(2016)067 

\bibitem{Calcagni}  Calcagni G and Modesto L,
{\it   Phys.\ Lett.} B {\bf 773}, 596 (2017)
  ({\it Preprint} gr-qc/1707.01119)
  
\bibitem{Modesto:2010uh} Modesto L,  Moffat J W and Nicolini P,
  {\it Phys.\ Lett.} B {\bf 695}, 397 (2011) ({\it Preprint}  gr-qc/1010.0680)

\bibitem{completeness} Bambi C, Modesto L and Rachwal L, 
{\it J. Cosmol. Astropart. Phys.} JCAP {\bf 1705}, no. 05, 003 (2017)

\bibitem{evaporation} Bambi C, Modesto L, Porey S and Rachwal L, 
{\it J. Cosmol. Astropart. Phys.} JCAP {\bf 1709}, 033 (2017)

\bibitem{entanglement} Giaccari S, Modesto L, Rachwal L and Zhu Y, 
{\it Preprint} hep-th/1512.06206

\end{thebibliography}
\end{document}